\documentstyle[emulateapj,psfig]{article}
\def\shortauthors{\lefthead}
\def\shorttitle{\righthead}

\newcount\levelone    \levelone=0
\newcount\leveltwo    \leveltwo=0
\newcount\levelthree  \levelthree=0
\newcount\levelfour   \levelfour=0
\def\chaphead{}                             
\def\secno{\chaphead\the\levelone}
\def\subno{\chaphead\the\levelone.\the\leveltwo}
\def\subsubno{\chaphead\the\levelone.\the\leveltwo.\the\levelthree}
\def\subsubsubno{\chaphead\the\levelone.\the\leveltwo.\the\levelthree
                           .\the\levelfour}
\def\newsec{\advance\levelone by1 \leveltwo=0 \levelthree=0 \levelfour=0}
\def\newsub{\advance\leveltwo by1 \levelthree=0 \levelfour=0}
\def\newsubsub{\advance\levelthree by1 \levelfour=0}
\def\newsubsubsub{\advance\levelfour by1}

\def\absnarrower{\advance\leftskip by \abstractindent}

\newdimen\secskipamount  \secskipamount=1pt
\newdimen\subskipamount  \subskipamount=1pt

\newdimen\bottomtol \bottomtol=0.03\vsize
\def\secskip{\par \ifdim\lastskip<\secskipamount \removelastskip \fi
    \vskip 0pt plus \bottomtol \penalty-250
    \vskip 0pt plus -\bottomtol \relax
    \vskip\secskipamount plus3pt minus3pt}
\def\subskip{\par \ifdim\lastskip<\subskipamount \removelastskip \fi
    \vskip 0pt plus 0.5\bottomtol \penalty-150
    \vskip 0pt plus -0.5\bottomtol \relax
    \vskip\subskipamount plus2pt minus2pt}
\def\subsubskip{\par \ifdim\lastskip<\subskipamount \removelastskip \fi
    \vskip 0pt plus 0.5\bottomtol \penalty-150
    \vskip 0pt plus -0.5\bottomtol \relax
    \vskip\subskipamount plus2pt minus2pt \hskip 10pt}

\outer\def\unnumberedsectionbegin #1 #2\par {\secskip \noindent {{\bf  #1}
\dotfill #2}
    \nobreak \vskip 1pt \noindent}

\outer\def\sectionbegin #1 #2\par {\secskip \newsec \noindent {{\bf \secno\  #1}
\dotfill #2}
    \nobreak \vskip 1pt \noindent}
\outer\def\subsectionbegin #1 #2\par {\subskip \newsub {\subno\ {\rm #1} \hfill #2}
    \nobreak \vskip 1pt \noindent}
\outer\def\subsubsectionbegin #1 #2\par {\subsubskip \newsubsub
    {\subsubno\ {\it #1} \hfill #2}
  \nobreak \vskip 1pt \noindent}

\newcount\eqnumber \eqnumber=0
\def\new{{\rm\chaphead\the\eqnumber}\global\advance\eqnumber by 1}

\newcount\fignumber \fignumber=0
\def\nfig{\chaphead\the\fignumber\global\advance\fignumber by 1}
\def\ntab{\chaphead\the\tabnumber\global\advance\tabnumber by 1}

\newcount\fononum \fononum=0
\def\nfn{\global\advance\fononum by 1}
\def\fonono{\the\fononum}

\def\bck{\hskip-0.35em}

\def\wisk#1{\ifmmode{#1}\else{$#1$}\fi}
\def\etal{{et al.$\,$}}

\def\msyr{\wisk{\,\rm M_\odot\,yr^{-1}}}
\def\cse{circumstellar envelope}
\def\cses{circumstellar envelopes}

\def\losn{line of sight}

\def\decdeg#1.#2 {\wisk{#1^{\,\rm o}\bck.\,#2}\ }
\def\decmin#1.#2 {\wisk{#1^{\,\prime}\bck.\,#2}\ }

\def\decsec#1.#2 {\wisk{#1^{\prime\prime}\hskip-0.42em.\hskip0.10em#2}\ }
\def\arcsec {\wisk{^{\prime\prime}}\ }
\def\kms{\wisk{\,\rm km\,s^{-1}\,}}                    

\def\oversim#1#2{\lower1.5pt\vbox{\baselineskip0pt \lineskip-0.5pt
     \ialign{$\mathsurround0pt #1\hfil##\hfil$\crcr#2\crcr\sim\crcr}}}

\def\gsim{\wisk{\mathrel{\mathpalette\oversim{>}}}} 
\def\lsim{\wisk{\mathrel{\mathpalette\oversim{<}}}} 

\def\apnii#1 { {2000, In: {Kastner J., Soker N., Rappaport S. (eds.)
     Asymmetric Planetary Nebulae II}. ASP Conf.Ser.199, Chelsea, p. #1}}

\def\physr#1 {, {\it Physics Report}{\bf#1},\ }
\def\iauc#1 #2 {, {IAU Circ., }{#1, #2}\ }
\def\nature#1 #2 {, {Nat \ }{#1, #2}\ }
\def\science#1 #2 {, {Sci \ }{#1, #2}\ }
\def\aa#1 #2 {, {A\&A,}{ #1, #2} }
\def\aal#1 #2 {, {A\&A,}{ #1, L#2}\ }
\def\aas#1 #2 {, {A\&AS,}{ #1, #2} }
\def\aj#1 #2 {, {AJ, }{#1, #2}\ }
\def\apj#1 #2 {, {ApJ, }{#1, #2}\ }
\def\apjl#1 #2 {, {ApJ, }{#1, L#2 }\ }
\def\apjs#1 #2 {, {ApJS, }{#1, #2}\ }
\def\araa#1 #2 {, {ARA\&A, }{#1, #2}\ }

\def\mnras#1 #2 {, {MNRAS, }{#1, #2}\ }

\def\rpphys#1 #2 {, {Rep. Prog. Phys., }{#1, #2}\ }

\def\pasp#1 #2{, {PASP, }{#1, #2}\ }
\def\qjras#1 {, {QJRAS, }{#1}\ }
\def\aus#1 #2{, {Aust.~J. Phys., }{#1, #2}\ }
\def\actaa#1 {, {Acta Astron., }{#1}\ }

\def\agbs#1 { {1999, In: {Le Bertre T., Lebre A., Waelkens C. (eds.)
     Proc. IAU Symp. 191,
     Asymptotic Giant Branch Stars}. ASP, Chelsea, p. #1 }}

\def\eqref#1{\advance\eqnumber by -#1 \chaphead\the\eqnumber
           \advance\eqnumber by #1 }
\def\?{\eqref{1}}
\def\last{\advance\eqnumber by -1 {\rm\chaphead\the\eqnumber}\advance
     \eqnumber by 1}
\def\eqnam#1{\xdef#1{\chaphead\the\eqnumber}}
\def\appendixbegin#1 #2{\eqnumber=1 \def\chaphead{{#1}}
    \levelone=0\leveltwo=0\levelthree=0\levelfour=0\eqnumber=1\fignumber=1 
    \vskip\subskipamount\noindent{\ninepoint\bf Appendix #1\ \ \ #2}
    \vskip\subskipamount\noindent}
\def\noappendixbegin#1 #2{\eqnumber=1 \def\chaphead{{#1} }
    \levelone=0\leveltwo=0\levelthree=0\levelfour=0\eqnumber=1\fignumber=1 
    \vskip\subskipamount\noindent{}
    \vskip\subskipamount\noindent}
\def\nfig{\chaphead\the\fignumber\global\advance\fignumber by 1}
\def\anfig{\global\advance\fignumber by 1}

\def\ntab{\chaphead\the\tabnumber\global\advance\tabnumber by 1}
\def\antab{\global\advance\tabnumber by 1}
\def\nfiga#1{\chaphead\the\fignumber{#1}\global\advance\fignumber by 1}
\def\rfig#1{\advance\fignumber by -#1 \chaphead\the\fignumber
            \advance\fignumber by #1}
\def\fignam#1{\xdef#1{\chaphead\the\fignumber}}
\def\tabnam#1{\xdef#1{\chaphead\the\tabnumber}}
\newcount\fignumber \fignumber=1
\newcount\eqnumber \eqnumber=1
\newcount\tabnumber \tabnumber=4

\def\Sct{Section\ }

\def\Fg{Figure\ }

\def\bibitm{\bibitem{}}
  
\def\sat{1612--MHz}
\def\sath{1720--MHz}
\def\mla{1665--MHz}
\def\mlb{1667--MHz}
\def\Btnt{OH009.1$-$0.4}
\def\ALL{1}
\def\TAIL{2}

\begin{document}

\title{A shock--excited OH maser in a post--AGB envelope ?}

\author{Maartje N.~Sevenster}
\affil{MSSSO/RSAA, Cotter Road, Weston ACT 2611, 
 Australia (msevenst@mso.anu.edu.au)}

\author{Jessica M.~Chapman}
\affil{ATNF, P.O.Box 76, Epping NSW 1710, Australia}

\shortauthors{M.~Sevenster, J.~Chapman}
\shorttitle{A shock--excited OH maser in a post--AGB envelope ?}

\begin{abstract}
We have observed a sample of OH 1612--MHz masing objects in all four
OH ground--state transitions with the Australia Telescope Compact Array.
One likely post--AGB object is found 
to emit in the \sat , \mla\ and \sath\ transitions. We
discuss the evidence that this object may be an
early post--AGB object and the possibility
for such a circumstellar envelope to harbour a \sath\ maser.
We argue that during a very brief period, just after the star 
has left the thermally--pulsing phase of the AGB and the 
wind velocity starts to increase, post--AGB objects
might show \sath\ emission. The best objects to search for such
emission would be those that are masing at 1612 MHz and 
1665 MHz, but not at
1667 MHz nor in the 22--GHz H$_2$O transition.
\end{abstract}


\section{Introduction}

OH maser emission from the four ground-state transitions at 1612.231,
1665.402, 1667.359 and 1720.530 MHz occurs in a wide range of objects.
Extra--galactic OH--maser emission is associated with active galactic
nuclei and enhanced central star formation (Baan \& Haschick 1987;
Randell \etal 1995), with strongest emission, in general, from the
main lines at 1665/1667 MHz. OH maser
emisson from galactic star forming regions (SFRs) is also strongest in
the main lines, with dominant emission mostly at 1665 MHz.

The satellite line at 1720 MHz is best known
for its occurence in supernova remnants, whereas that
at 1612 MHz is strongest in the optically--thick envelopes of 
oxygen--rich asymptotic--giant--branch (AGB) stars. 
Mainline OH emission may be present in optically--thin envelopes; from
nearby Mira variables, with low mass-loss rates around 10$^{-7}$ M$_\odot$
yr$^{-1}$, only mainline emission is detected.

At the end of the AGB evolutionary phase, the star undergoes a
transformation which takes a couple of thousand years, during which it
evolves to become a planetary nebula (PN). During this post-AGB stage,
the mass-loss rate decreases by a factor of 1000 or more whilst the
stellar surface temperature rises by a factor of 10 or 20 to above
50,000 K.  The star changes from losing mass in a cool, dense wind to
losing mass in a much hotter, lower density wind. 

Strong changes also
occur in the envelope kinematics and morphologies. In particular,
whilst the circumstellar envelopes of AGB stars are generally
spherically symmetric, images of planetary nebulae show axi-symmetric
morphologies in over 50\% of cases (Manchado \etal2000). 

During the initial searches in the 1970s
it was quickly found that \sath\ emission was not detected from AGB
stars. As this agreed with theoretical expectations 
(eg.~Elitzur, Goldreich \& Scoville 1976), there have been
no systematic searches for this maser in stellar sources.
However, it is imaginable that the shocks in some
post--AGB stars could temporarily induce \sath\ maser emission.
We therefore decided to observe, for the first time, a modest sample of
\sat --selected sources, including post--AGB objects, at 1720 MHz. A
total of 18 objects were selected from Sevenster \etal(1997a,b) on the
basis of their unusual OH 1612--MHz profiles.
Observations in all four ground--state 
OH transitions were taken with the Australia Telescope Compact Array (ATCA)
in November 1998 (1612, 1665 and 1667 MHz) and September 1999 (1720
MHz). Eleven of the sources are likely post--AGB objects.

The full results from this study will be published separately. Here we
report the detection of 1720 MHz from the source \Btnt\ (IRAS 18043$-$2116).
In \Sct 2, we present the
observations and in \Sct 3 the evidence that \Btnt\ is a post--AGB
object. In \Sct 4 we discuss a possible scenario for the \sath\ excitation. 
Conclusions are given in \Sct 5.

\section{Compact Array results}

Figure \ALL\ shows the OH spectra for \Btnt\ (18:07:20.9,
$-$21:16:10.9 (J2000)), observed with a channel width of 0.7 \kms.
The maser emission is strongest at 1612 MHz, with a peak flux density
of 11.0 Jy at 71.2 \kms.  The spectral profile has a `one-sided' peak
and a much weaker feature at 101 \kms.  The \mla\ emission covers the
same velocity range as the \sat\ emission (\Fg \TAIL ), with a peak
flux density of 1.64 Jy at 71.3 \kms.  No emission was detected at
1667 MHz with a detection limit of $\sim$ 80 mJy.

The \sath\ emission is seen as a single narrow peak at a slightly more
blue-shifted velocity than the \sat\ and the \mla\ peaks. The peak
flux density is 1.22 Jy at 69.4 \kms , with no linear polarization.
The positional coincidence (\decsec 0.3) of the emission in all three
lines is consistent with a single point source, given 
the ATCA resolution ($\sim$3\arcsec).

We also searched for radio--continuum emission at 4.8 and 8.6 GHz with
no detection at the source position, 
resulting in an upper flux limit of 1.5 mJy (five-$\sigma$).

\begin{figure*}
\anfig
\psfig{file=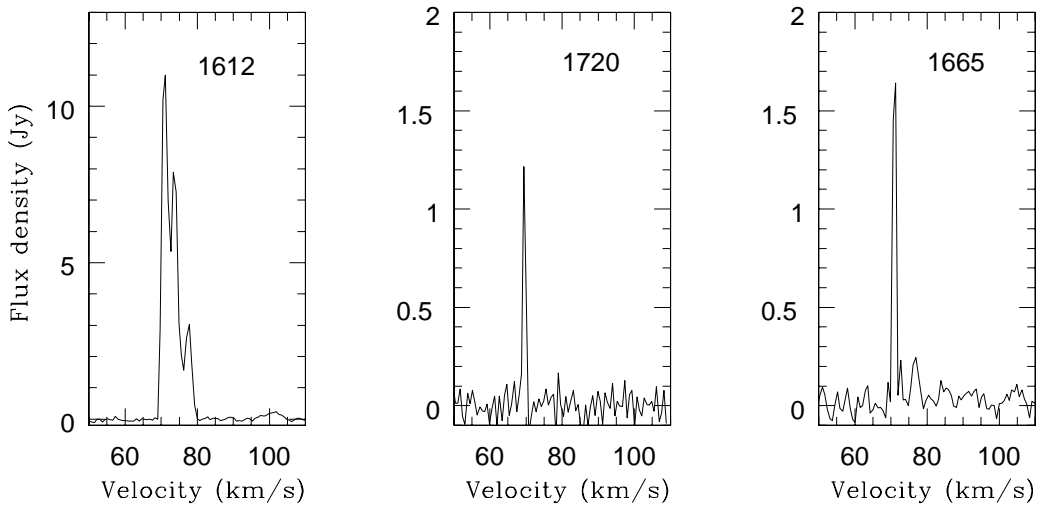,width=15truecm}
\vskip -6.5truecm
\figcaption{
The \sat , \sath\ and \mla\ spectra
for \Btnt , as observed in November 1998 (1612,1665 MHz) 
and September 1999 (1720 MHz). 
The stellar velocity is likely to be around 85 \kms .
The \sath\ peak is slightly offset from the \sat\ and \mla\ peak 
velocities (see \Sct 3). All spectra were observed with a channel
width of 0.7 \kms .
}
\end{figure*}

\section{The evolutionary status of \Btnt }

We consider it extremely unlikely that \Btnt\ is associated
with either a SFR or a supernova remnant. No known supernova remnant is
associated with the \sath\ position and the presence of the other
two OH masers precludes such a classification (eg.~Koralesky \etal1998).

The non--detection of radio continuum is unfavourable to a
classification as a SFR, since most SFRs with OH masers are also strong
(\gsim 100mJy) ultra--compact HII regions (Caswell 1998).
Furthermore, a non--detection in methanol at 6668 MHz at the position
of \Btnt\ (Jim Caswell, private communication) provides an upper limit
for methanol--maser emission at which 90\% of SFRs would be detected
if they show \sath\ maser emission (Caswell 1999).

The far-infrared MSX and IRAS colours of $R_{21}$=$\log
(S_{25}/S_{12})$=0.53 and $R_{32}$=$\log
(S_{60}/S_{25})$=0.4\footnote{$S_{12}$ is taken from the Midcourse
Space Experiment infrared satellite (MSX), see
http://www.ipac.caltech.edu/ipac/msx/msx.html. $S_{25}$ and $S_{60}$
are IRAS point--source values.} are consistent with a classification as
a ``high--outflow'' source (Zijlstra \etal2000). These sources include
both young stellar objects and bipolar post--AGB stars with optically
thick envelopes. In the two--colour diagram presented by Pottasch
\etal(1988), of the eight nearest neighbours to \Btnt , five are PNe,
one is an OH/IR star and two are galaxies. For \Btnt , the
$R_{32}$ is bluer than for typical HII regions and far redder
than for Mira variables. The high $R_{21}$ indicates a
(recent) mass--loss rate of $>\,10^{-5}$\msyr\ (Van der Veen \etal
1995), consistent with an AGB or post--AGB classification.

The OH spectral profiles strongly suggest that the source is a young
post--AGB star. The steep outer edge of the bright \sat\ peak is
characteristic of OH/IR stars and indicates that the emission arises
in a thin, expanding shell.  The raggedness on the inside of the peak
suggests that the star is transiting off the AGB, with irregular or
intermittent mass loss.  A very large flux--density ratio between the
blue-- and the red--shifted peak has previously been seen in post--AGB
stars, where the red--shifted emission, from the back of the \cse , is
absorbed in the small ionized region around the star, as noted by
Lewis (1989). A similar spectrum is found for OHPN 9 (Zijlstra \etal 1989), 
which shows radio--continuum emission of 1.4 mJy, that could have
gone undetected with our limit of 1.5 mJy.
Approximately 50\% of PNe have
radio--continuum emission weaker than this (e.g.\ Pottasch \etal 1988).
For an object with an optical depth of one, a $T_{\rm eff}$ of 3000 K,
a distance of 5 kpc and an ionization radius of $10^{15}$cm, we
estimate that the 6--cm flux density would be $\sim$ 1 mJy.

Also consistent with a classification as a young post--AGB star are
the non--variability of the IRAS and MSX emission from \Btnt\ and the
detection of mainline OH emission that is weaker than the \sat\
emission (Lewis 1989). Here we suggest that the absence of \mlb\
emission may be a signature of the earliest stages of post--AGB
evolution.  In \cses , the \mla\ emission can be stronger than that at
1667 MHz in regions with dust temperatures below 300 K, if the
far--infrared spectral index is \gsim 2 (Elitzur 1978).  Pavlakis \&
Kylafis (1996) find that, at lower temperatures, the \mla\ emission
can be stronger also for spectral indices of $\sim$1. For envelopes which
have only dilated enough for ambient H$_2$O--dissociating photons to
reach regions with $T_{\rm d}<$300 K, the \mlb\ maser emission is
expected to be weak or absent. With further dilation, the \mlb\ emission
should increase and the \sat\ emission weaken.

From the infrared and OH maser properties and the non-detections of
methanol--maser and continuum emission, we conclude that \Btnt\ is
almost certainly a young post--AGB star.

\section{Discussion}

It is very unlikely that the \sath\ emission originates from exactly
the same regions as the other lines (see 
Lockett, Gauthier \& Elitzur 1999). In the few SFRs that have been
imaged in both satellite masers, they come from spatially separate
clumps (Caswell 1999).
However, the two satellite transitions require
conditions similar enough to compete directly, as is seen very clearly
by their conjugate behaviour in Centaurus A (van Langevelde \etal
1995).
During the transition from the AGB to the PN phase, temperatures,
densities, wind velocity and irradiation change rapidly by orders of
magnitude. A variety of environments must be present in
the \cse\ and it is possible that locally conditions could
favour the \sath\ over the \sat\ maser.

We hypothesize that the \sath\ transition is 
collisionally excited in a region where shocks are
present, at the interaction region between
the remnant of the AGB superwind phase and the
hotter, fast post--AGB wind. 
Interacting--wind models usually assume a spherical
fast wind colliding with an equatorially--enhanced density
distribution of the AGB remnant; another possibility is
that a bipolar fast wind pierces into a spherical AGB remnant.
Indeed, Weintraub \etal(1998) argue that it is likely
that shocks occur after the onset of bipolarity and
before the nebular envelope is ionized and the masers are destroyed.

A C--type of shock
is the only way to create the right conditions for the \sath\ maser to
exist (Lockett \etal 1999).
The process could be as follows. In the very early post--AGB
evolution (see Sch\"onberner \& Steffen 2000), 
the wind picks up in velocity to $\sim$ 100 \kms, while the mass--loss
rate drops significantly. In the first $\sim$1000 years, a
shock front of $\sim$50 \kms\ would travel $\sim$ 2$\times$10$^{17}$
cm (see Frank \& Mellema 1994, their Fig.2: "swept--up shell";
cf.~Bujarrabal \etal1997, M1$-$92).  
This is well into the regions of the AGB envelope where the
external radiation field dissociates and ionizes most molecular
material ($n$\lsim $10^4$cm$^{-3}$, $T$\lsim 100 K).  (Note
that 2$\times$10$^{17}$ cm is not resolved by ATCA at distances beyond
5 kpc.)

The passage of a C--type shock would change the chemistry in this
region, forming H$_2$O and increasing the density by a factor of 10
(see Wardle 1999). Subsequent dissociation of the water by the ambient radiation
field sufficiently enhances the OH abundance, while collisions with
preferentially ortho--H$_2$ (see Lockett \etal 1999)
leave OH in the proper excited state to decay via the \sath\
transition, provided cooling has also decreased the post--shock
temperature to below 125 K (Lockett \etal 1999).  
With an ionizing rate of
$\sim$10$^{-17}$s$^{-1}$ (solar vicinity 2$\times$10$^{-17}$s$^{-1}$,
IRC+10216 2$\times$10$^{-18}$s$^{-1}$, Glassgold 1999) and other
parameters as given before, it is feasible that the necessary column
density of $>$2.5$\times$10$^{15}$ cm$^{-2}$ can be achieved (Lockett
\etal 1999; Wardle 1999).  For this, we assume the relevant layer has
a thickness of $\sim$10$^{17}$ cm (the outer radius of the \sat\ masing
shell of an OH/IR star is typically 10$^{17}$ cm, Cohen 1989), the
post--shock density is \lsim 10$^5$ cm$^{-3}$ with OH abundance 
of 10$^{-5.5}$ (in the \sat\ layer this is $\sim$10$^{-4}$, Cohen 1989) and the
magnetic field at these radii is of the order of 100$\mu$G
(e.g.~Chapman \& Cohen 1986). These values are realistic, as well as in 
good agreement with those used by Wardle (1999).  Note that the 
resulting \sath\ maser optical depth would
be low, of the order of 4 (Lockett \etal 1999).

In this scenario, the \sath\ masing regions are located
well outside the layers where the \mla\ masers arise and along a different
\losn\ than the \sat\ masers.
This means that the usual argument that the \sath\ maser cannot co--exist 
with these lines in one circumstellar stellar envelope (see footnote 
Lockett \etal 1999) is not necessarily valid.
The formation of extra water at large radii is not in disagreement with
the observation that post--AGB objects with main--line emission do not
have 22--GHz water masers (Lewis 1989) as no maser emission is expected
from the shock--produced H$_2$O (Lockett \etal 1999).

\anfig
\vskip 0.2truecm
\psfig{file=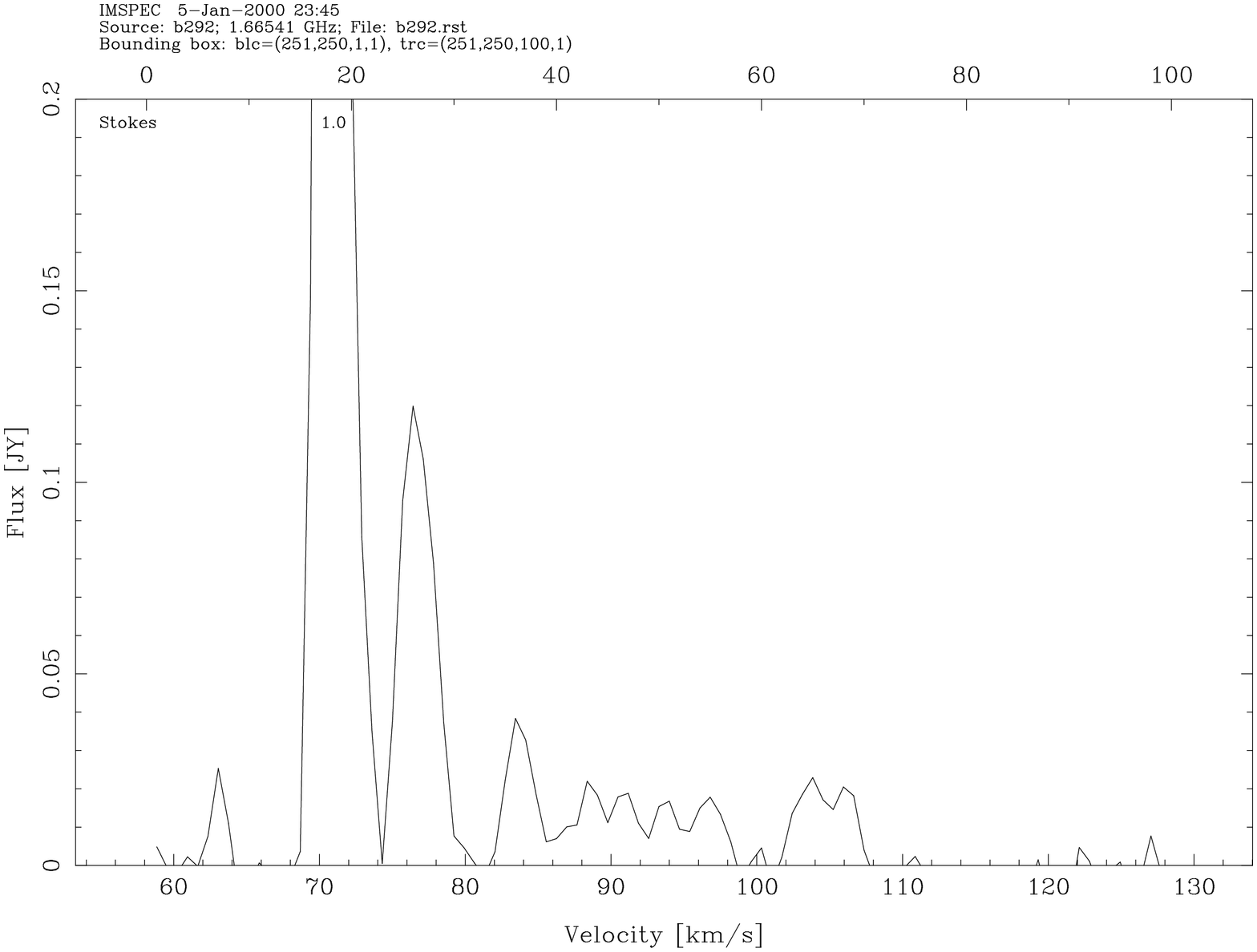,angle=270,width=8truecm}
\figcaption{
The hanning--smoothed cleaned--image spectrum at 1665 MHz, 
at the position of the star (maximum pixel). The \mla\ emission
is seen over the same range of velocities as 1612 MHz, but 
the detection is only significant around 72 \kms .
}
\vskip 1truecm

The high--density ($>$10$^7$) regime for \sath\ maser emission, as presented
in Pavlakis \& Kylafis (1996), requires strong 
velocity gradients ($\sim$2\kms ; see Caswell 1999 for such spectra). 
As the linewidth for the \sath\ maser is \lsim 0.5\kms ((\Fg\ALL ),
typical for velocity gradients in \cses , such a regime appears to be unlikely.

The detection probability for \sath\ maser emission from post--AGB
stars is almost certainly very low for several
reasons. Firstly, a C--type shock has to form.  Because of the
geometry of the shock front and maser beaming effects, it is likely
that only around 10\% of sources with such shocks will show 
\sath\ emission, as is found for supernova remnants (Lockett \etal 1999;
Koralesky \etal 1998; Claussen \etal 1997).  In addition, the
interstellar radiation field has to be strong enough, but not so
strong as to increase the electron density (Lockett \etal 1999).
Finally, in many objects a bipolar flow is traced by the \sat\ line
(eg.~Zijlstra \etal 2000, Sahai \etal1999). In such cases, the physical
conditions in the interaction region 
cannot be suitable for \sath\ emission, probably due to large
column density or the strength 
of the local infrared radiation field (Lockett \etal1999).

Circumstellar \sath\ maser emission was previously detected from the
star V1057 Cygni (Lo \& Bechis 1973; Winnberg \etal 1981), following
an optical/infrared flare in 1970. It has been suggested that the
\sath\ flares occured when shells ejected at different velocities
collided with remnant star--forming material and that the masers were
collisionally excited (Elitzur 1976). The maser flares only lasted for
$\sim$ 18 months, showing that the right conditions for \sath\ maser
emission may be very short--lived.

\subsection{Predictions}

Long--baseline interferometry of \Btnt\ will be carried out with MERLIN
in the near future and may help establish whether it is
the AGB superwind or the proto--planetary wind that is aspherical and causes 
the wealth of morphologies observed in PNe.
In our scenario, in the former case, the \sath\ masers 
would be preferentially on the equator of the OH shell,
in the latter on its poles. We predict that this is more likely,
given the very narrow width of the \sath\ line, compared 
to the \sat\ profile. On average, the \sath\ maser will be
located outside the \sat\ and \mla\ layers, which is in agreement with its
slightly higher velocity in an accelerating outflow (see Bujarrabal \etal1997).
The absence of the red--shifted peak at 1720 MHz suggests that
we see the bipolar outflow nearly end--on (cf.~Claussen \etal1997).

If the object is a post--AGB star,
measuring Zeeman splitting of the narrow \sath\ lines
would give a measurement of the magnetic field, as has been done
for several supernova remnants (eg.~Brogan \etal2000; Koralesky \etal1998), and provide
interesting boundary conditions for magneto--hydro--dynamic
models (see Garcia--Segura, Franco \& Lopez 2000).

Finally, we predict that no H$_2$O--maser emission will be found
in this object, but that thermal emission should be detectable given
the large column density.
Several H$_2$O transitions in the far infrared (eg.~179.5$\mu$m,174.5$\mu$m)
could provide conclusive evidence in favour of the shock--formed water scenario
(Kaufman\& Neufeld 1996), but other shock indicators, such as near--infrared
H$_2$ transitions may be more practical for observations.

\section{Conclusions}

We have detected an OH \sath\ maser in a likely post--AGB
envelope. From theoretical as well as observational arguments,
it is possible that this maser exists briefly in such environments,
at least in some sources.
We argue that the \sath\ maser could arise in the
region where the AGB-- and the starting PN winds collide.
We expect that this is a short-lived event
and that our detection rate of 1/11 is due to favourable small number
statistics. The most likely objects to show this
emission may be those that have fairly strong emission
at 1612 MHz, with weaker \mla\ and possibly SiO maser emission,
but no \mlb\ and certainly no H$_2$O 22--GHz maser emission.

\begin{acknowledgements}
We thank Mark Wardle and Jim Caswell for very helpful discussions on
this work. The Australia Telescope Compact Array is operated as part
of the Australia Telescope National Facility, which is funded by the
Comnmonwealth of Australia for operation as a National Facility
managed by CSIRO.
\end{acknowledgements}

\end{document}